\documentclass[twocolumn]{aastex631}
\usepackage{amsmath}
\usepackage{ulem}

\begin{document}
\title{The Role of a Neutron Component in the Photospheric Emission of Long-Duration Gamma-Ray Burst Jets}

\author[0009-0008-1678-2787]{Nathan Walker}
\affiliation{Oregon State University \\
Department of Physics, 301 Weniger Hall, Oregon State University \\
Corvallis, OR 97331, USA}
\author[0000-0002-4299-2517]{Tyler Parsotan}
\affiliation{Astrophysics Science Division, NASA Goddard Space Flight Center,Greenbelt, MD 20771, USA}
\author[0000-0002-9190-662X]{Davide Lazzati}
\affiliation{Oregon State University \\
Department of Physics, 301 Weniger Hall, Oregon State University \\
Corvallis, OR 97331, USA}

\begin{abstract}

Long-duration gamma-ray bursts (LGRBs), thought to be produced during core-collapse supernov\ae, may have a prominent neutron component in the outflow material.  If present, neutrons can change how photons scatter in the outflow by reducing its opacity, thereby allowing the photons to decouple sooner than if there were no neutrons present.  Understanding the details of this process could therefore allow us to probe the central engine of LGRBs, which is otherwise hidden.  Here, we present results of the photospheric emission from an LGRB jet, using a combination of relativistic hydrodynamic simulations and radiative transfer post-processing using the Monte Carlo Radiation Transfer (MCRaT) code. We control the size of the neutron component in the jet material by varying the equilibrium electron fraction $Y_{e}$, and we find that the presence of neutrons in the GRB fireball affects the Band parameters $\alpha$ and $E_{0}$, while the picture with the $\beta$ parameter is less clear.  In particular, the break energy $E_{0}$ is shifted to higher energies. Additionally, we find that increasing the size of the neutron component also increases the total radiated energy of the outflow across multiple viewing angles. Our results not only shed light on LGRBs, but are also relevant to short-duration gamma-ray bursts associated with binary neutron star mergers, due to the likelihood of a prominent neutron component in such systems.

\end{abstract}



\section{Introduction}
Our understanding of Gamma-Ray Bursts (GRBs) has evolved dramatically since their discovery in the late 1960's. First detected as short transient bursts of high energy photons \citep{klebesadel_observations_1973}, observations of afterglows \citep{groot_search_1998, costa_discovery_1997} and supernova counterparts \citep{galama_unusual_1998,hjorth_very_2003,woosley_supernovagamma-ray_2006,bloom_unusual_1999} have facilitated a deeper understanding of these otherwise mysterious events. 
Long duration gamma-ray bursts (LGRBs) are now thought to occur during core-collapse supernov\ae, a process in which stars more massive than about $8 M_{\odot}$ end their lives in a violent explosion, resulting in the formation of either a Black Hole (BH) or a Neutron Star (NS) \citep{woosley_physics_2005}. After the formation of either a BH or a NS, material from the preceding collapse can accrete around the compact object, providing a possible power source for an ensuing LGRB (e.g. \cite{narayan_accretion_2001}). Alternatively, a highly magnetized, fast spinning NS could power a relativistic outflow by tapping into its rotational energy (e.g., \citealt{Bucciantini2012}).  Given the possibility of a NS as either an intermediate or a terminal stage of the supernova, there is a strong possibility of a neutron component in the accreting material, which can then be collimated into a  relativistic jet and produce a LGRB. 

In spite of this progress, one aspect of GRBs that still remains in contention is the nature of the prompt emission. In LGRBs, the prompt emission can last anywhere from a few seconds to a few minutes \citep{bloom_unusual_1999,macfadyen_supernovae_2001} and is characterized by bright, non-thermal spectra \citep{band_batse_1993}.  A leading model that explains this emission is the Synchrotron Shock Model (SSM). In this model, the jet expands and reaches the photosphere without producing noticeable radiation. After passing the photosphere, electrons in colliding internal shocks produce non-thermal radiation \citep{rees_unsteady_1994}.  While this model naturally explains the characteristic non-thermal emission of GRBs and is able to fit the spectra of a number of bursts, it may have difficulties in reproducing the peak width of bursts \citep{beloborodov_regulation_2013}. In addition, some burst have spectra that are inconsistent with a simple model in which electrons are accelerated impulsively and either do not cool (the line-of-death problem, \citealt{preece_synchrotron_1998}) or cool radiatively  \citep{ghisellini_constraints_2000}. Finally, the SSM model has difficulty reproducing the ensamble correlations between properties of different bursts, such as the Amati and the Yonetoku correlations \citep{amati_intrinsic_2002,yonetoku_gamma-ray_2004}.

A viable alternative to the SSM is the so-called photospheric model (e.g. \cite{beloborodov_collisional_2010-1}, \cite{giannios_spectral_2007}, \cite{lazzati_very_2009}, \cite{ryde_observational_2011}, \cite{peer_observable_2006}). In this model, thermal radiation is produced when the jet is hot and dense near the central engine. As the jet propagates and expands the radiation is shaped through its interaction with the expanding outflow. Effects such as  sub-photospheric dissipation \citep{chhotray_gamma-ray_2015,parsotan_photospheric_2018,ito_monte_2018} and multi-color blackbody emission \citep{peer_theory_2011} enable this model to account for a non-thermal spectrum. Additionally, as the radiation scatters and propagates with the outflow, it is imprinted with a signature of the history of the outflow that survives until the radiation escapes at the photosphere \citep{vurm_radiative_2016}.  Because of this, the composition and dynamics of the jet material are of crucial importance in shaping the observed prompt emission in the photospheric model.  An important test of the photospheric model can then be to model the effect that different compositions of the jet material can have on radiation.

Given the possibility of a neutron component in both the compact mergers and supernov\ae{} that are thought to produce GRBs, a body of work has been produced that explores the consequences of a neutron component in GRB fireballs. This includes a detailed study on the processes that shape the nuclear composition of the fireball as it expands \citep{beloborodov_nuclear_2003}, the role neutrons play in heating the jet through collisional processes \citep{beloborodov_collisional_2010,rossi_heating_2004}, and that of the dynamics of shocks in the explosion \citep{derishev_neutron_1999}.  However, no work has been done on how neutrons directly shape the observed prompt emission of GRBs. Therefore the role of a neutron component on the photosepheric emission of a LGRB is of particular interest, and a good candidate to further test the photospheric emission model. 

In this paper, we use the MCRaT radiative transfer code and the ProcessMCRaT python package \citep{lazzati_monte_2016,parsotan_monte_2018,parsotan_photospheric_2018,parsotan_photospheric_2021} to scatter photons through a 2D relativistic hydrodynamic (RHD) simulation of a LGRB jet \citep{morsony2007,lazzati_photospheric_2013}, and produce mock observables. We control the relative size of the neutron component in the jet material by varying the equilibrium proton-to-nucleon ration $Y_{e}$. This paper is organized as follows: in Section 2 we summarize how the MCRaT code scatters photons and describe how we take into account a neutron component in the jet; in Section 3 we present results of spectra obtained by varying $Y_{e}$; and in Section 4 we discuss our results and their implications.


\section{Computational Methods} 
\subsection{The MCRaT Code}
We use the MCRaT radiative transfer code to individually Compton-scatter a set of photons injected into a RHD simulation of a LGRB jet. In this section we summarize the MCRaT algorithm. Further details on the original algorithm can be found in \cite{lazzati_monte_2016}, with improvements found in \cite{parsotan_monte_2018}.

MCRaT begins by injecting photons into the output of a RHD simulation.  During this injection process, MCRaT selects which RHD cells to inject photons into based on a set of user-defined parameters: the injection radius $R_{inj}$ and the angular interval $\delta \theta$, defined with respect to the jet axis.  All cells within the interval $\delta \theta$ and with a radius between $R_{inj} \pm \frac{1}{2} \, c \delta t$ are selected, where $c$ is the speed of light and $\delta t$ is the time interval of the selected RHD frame. Once the injection frames are selected, MCRaT determines the four-momentum of the injected photons in each cell by sampling a thermal distribution centered at the local co-moving temperature, 
\begin{equation}
    T^{'}_{i} = \left( 
    \frac{3p_i}{a} 
    \right)^{\frac{1}{4}},
    \label{eq:temp}
\end{equation}
where $p_i$ is the pressure of the fluid and $a$ is the radiation density constant. The injected photons are then weighted according to \citep{parsotan_photospheric_2018}
\begin{equation}
    dN_{i} =\frac{\xi \, T_{i}^{'3} \, \Gamma_{i}} {w} \, dV_{i},
    \label{eq:weight}
\end{equation}
where $dN_{i}$ is the expected number of photons in the $i^{th}$ RHD cell, $\xi$ is the photon number density coefficient  from $n_{\gamma} = \xi T^3$ ($\xi = 20.29$ for a Planck spectrum and $\xi = 8.44$ for a Wein spectrum), $T_{i}^{'}$ is the comoving fluid temperature, $\Gamma_i$ is the bulk Lorentz factor, $dV_{i}$ is the volume element of the RHD cell, and $w$ is the weight of the injected photons. MCRaT calculates the expected number of photons in each cell via Equation \ref{eq:weight}, and draws a photon number from a Poisson distribution with a mean given by the expected number of photons. MCRaT then sums over the photon numbers in each cell, and if the total number of injected photons so obtained lies outside the user-specified range, the weights are adjusted and the process of calculating the expected number of photons via Equation \ref{eq:weight} repeats. The final weights are those that result in a total number of injected photons that lies within the user-defined range. 

Once the injected photon properties are determined,  MCRaT scatters each photon according to the properties of the RHD simulation. To begin with, each photon is assigned a mean-free path according to \cite{abramowicz_appearance_1991}
\begin{equation}
    \lambda_i = \frac{dr}{d \tau_{T}} = 
    \frac{1}{\sigma_{T} \, n_{i}^{'} \, \Gamma_{i} \left( 1 - \beta_{i} \cos \theta_{fl,i} \right)},
    \label{eq:lambda}
\end{equation}
where $\sigma_T$ is the Thomson cross section, $n_{i}^{'}$ is the comoving lepton number density, $\beta_i$ is the fluid velocity in units of $c$, and $\theta_{fl,i}$ is the angle between the fluid velocity and the photon velocity.  A random scattering time for each photon is drawn from the distribution
\begin{equation}
    P_{i}(t) \propto e^{-\frac{c}{\lambda_{i}} \, t},
    \label{eq:scatterTime}
\end{equation}
 and if the smallest of those scattering times is within the time interval of the given hydrodynamical simulation frame, the positions of the photons are all updated by allowing them to travel at the speed of light for the smallest scattering time obtained via Equation \ref{eq:scatterTime}.  Once all photons are updated to a new position in a frame, the photon with the shortest scattering time is scattered with an electron drawn from either a Maxwell-Boltzmann or a Maxwell-J{\"u}ttner distribution at the local fluid temperature, with a direction drawn from a random distribution.  If the smallest scattering times obtained from equation \ref{eq:scatterTime} lies outside the given RHD frame time interval, MCRaT allows the photons to propagate at the speed of light, without scattering, for an amount of time equal to the remaining time in the current RHD frame.  Then, MCRaT loads a new simulation frame and the photon mean free paths are all calculated again. This process of calculating photon properties and scattering with electrons is repeated for all the injected photons as they propagate and scatter through all of the provided RHD simulation frames.  
 
\subsection{Mock Observations}
When all the injected photons have been diffused beyond the photosphere we use the ProcessMCRaT package \citep{parsotan_photospheric_2022} to conduct mock observations.  This software allows for the injected photons to continue propagating unimpeded out to a virtual detector placed at a user-defined radius. To mimic a real observation in which the viewing geometry is unique we count only photons within a given acceptance range around the angle to the observer.   The energies of photons are obtained from the time component of the four-momentum at the end of the simulation, and the detection time is calculated as 
\begin{equation}
    t_{d} = t_{p} + t_{real} - t_{j}
    \label{eq:time},
\end{equation}
where $t_{real}$ is the simulation time at the frame used for an observation, $t_p$ is the photon detection time, and $t_j = r_{d} /c$ is the time it takes for a photon that was emitted at the instant the jet was launched to propagate to the detector. 

In the following, all light curves and spectra are obtained by placing the virtual detector at a radius of $r_d = 2.5 \times 10^{13} $ cm, which corresponds to approximately the edge of the RHD simulation.  When the photons haven't yet reached the last frame we find the positions of all the photons at the corresponding RHD simulation time and place a detector slightly beyond that point. 

After conducting a mock observation, we can bin the photon arrival times to calculate light curves,
\begin{equation}
    L_{t} = \frac{1}{\Delta \Omega \Delta t_{bin}} \sum_{i} w_{i} E_{i},
    \label{eq:lightcurve}
\end{equation}
where $E_{i}$ is the energy of each photon, $\Delta t_{bin}$ is the time bin, and $\Delta \Omega = 2\pi [\cos(\theta_{v} - \Delta \theta /2) - \cos(\theta_{v} + \Delta \theta /2)]$ is the solid angle the detector occupies, with $\Delta \theta$ being the angular acceptance range centered around $\theta_v$. We also bin the photon energies to calculate spectra via
\begin{equation}
    \frac{dN_{e}(E)}{dE \, dt} = \frac{1}{\Delta E_{bin} \Delta \Omega \Delta t} \sum_{i} w_{i},
    \label{eq:spectra}
\end{equation}
 where all the terms are the same as in Equation \ref{eq:lightcurve} except $\Delta E_{bin}$ and $\Delta t$, which are the energy bin width and the time interval over which photons were detected, respectively.

We fit the Band function \citep{band_batse_1993},

\begin{align}
\begin{split}
    &N_{E}(E) = \\   &  A \left( \frac{E}{100 \text{keV}} \right)^{\alpha} \exp(-E/E_{0}), 
                \\ & \qquad \qquad  \qquad \qquad \qquad E \leq (\alpha-\beta)E_{0} 
                \\ & A\left[ \frac{(\alpha-\beta)E_{0}}{100 \text{keV}} \right]^{\alpha-\beta} \exp(\beta - \alpha) \left( \frac{E}{100 \text{keV}} \right)^{\beta}
                \\ & \qquad \qquad  \qquad \qquad \qquad E \geq (\alpha-\beta)E_{0}.
                \label{eq:band}
\end{split}
\end{align}
to spectra obtained from equation \ref{eq:spectra}. In equation \ref{eq:band}, $\alpha$ and $\beta$ are the low and high energy slopes, respectively, $E_{0}$ is the break energy, and $A$ is related to normalization. The spectral peak is defined with respect to the spectral parameters in equation \ref{eq:band} as $E_{pk} = (2+\alpha) \, E_{0}$.

In order for the calculated spectra and light curves to correspond to what an observer would see, the optical depth must reach a value $\tau \sim 1$.  We calculate the optical depth \citep{parsotan_photospheric_2018} as:
\begin{equation}
    \tau_{i}^{n} = \sum_{j=i}^{L} \left< N \right>_{j}^{n} ,
    \label{eq:tau}
\end{equation}
where $L$ is the last frame of the RHD simulation and $n$ refers to a group of photons located initially in the $i^{th}$ frame, at some average position $R_{i}$. The sum over the RHD frame number $j$ goes from the $i^{th}$ frame to the last, with $\left< N \right>_{j}^{n}$ being the average number of scatterings that the $n^{th}$ group of photons experienced in the $j^{th}$ frame.  Equation (\ref{eq:tau}) essentially counts the number of scatterings each photon undergoes starting from the $i^{th}$ frame and we calculate it by tracking the number of scatterings that individual photons undergo, starting immediately after they are injected.  We similarly calculate the average energy of individual photons by tracking their energy throughout the MCRaT simulation. 

A group of photons is uncoupled from the jet if the average number of scatterings per photon starting from the $i^{th}$ RHD frame is $\lesssim 1$, corresponding to the photosphere condition of $\tau \sim 1$. Since this is computed separately for separate groups of photons it allows for the fact that photons in different parts of the jet and cocoon may uncouple at different times.

\subsection{A Neutron Component in the Fireball}
The MCRaT code reads in hydrodynamical data and determines the energy of injected photons via the hydrodynamical pressure (Equation \ref{eq:temp}), and their mean free paths via the hydrodynamical density and velocity (Equation \ref{eq:lambda}). Normally it is assumed that the total mass of the hydrodynamical simulation is attributed to protons (with a negligible contribution by electrons), and the lepton number density is therefore calculated by dividing the hydrodynamical density by the mass of the proton.  This picture changes when we include neutrons in our radiative transfer simulations. 

To simulate the role of a neutron component in the fireball, we use the  proton-to-nucleon ratio, $Y_e$, defined through the charge neutrality condition \citep{beloborodov_nuclear_2003}
\begin{equation}
    n_{-} - n_{+} = Y_{e} \, \frac{\rho}{m_{p}},
    \label{eq:charge}
\end{equation}
where $n_{\pm}$ are the $e^{\pm}$ number densities.  In the absence of $e^{\pm}$ pairs, $Y_{e}$ is just the electron-to-nucleon ratio and describes how many electrons there must be in a plasma in order to preserve charge neutrality. The density $\rho$ in Equation \ref{eq:charge} can in general consist of both protons and neutrons, and when both are taken into account the result is the equilibrium electron fraction $Y_{e} = n_{p} / (n_{p} + n_{n})$. Therefore, increasing the fraction of neutrons in the fireball decreases the electron-to-nucleon ratio, which in turn leaves fewer electrons with which to scatter photons.  When calculating photon mean free paths via Equation \ref{eq:lambda}, we can then scale the lepton density by $Y_{e}$.  A larger neutron component reduces the lepton density  of the jet.

A neutron component can in principle also change the hydrodynamical behavior of the plasma.  When the jet is still near the central engine it is hot and dense enough that the charged current reactions,
\begin{equation}
    e^{-} + p \rightarrow n + \nu, \, \, \, \, \, e^{+} + n \rightarrow p + \bar{\nu},
\end{equation}
establish an equilibrium $Y_{e}$. While these conditions will change as the jet expands, it has been shown that, further from the central engine, neutrons and ions can stay coupled through the acceleration stage as long as the jet has relatively high baryon loading \citep{beloborodov_nuclear_2003}.  In the same work it was also found that fireballs from neutron rich central engines are likely to remain neutron rich. We therefore do not consider the hydrodynamical effects of neutrons decoupling from protons, and we likewise keep the value of $Y_{e}$ constant throughout our MCRaT simulations. Since the baryons are treated as being in equilibrium we leave the pressure and velocity variables from the RHD simulation unchanged, and we scale the fluid density by the equilibrium electron fraction $Y_{e}$: $\rho \rightarrow Y_{e} \, \rho$.  

While we use a constant value of $Y_{e}$ for each MCRaT simulation we run, the RHD simulation is in fact comprised of material ejected from the central engine, a stellar envelope through which the jet must escape, and a radial power law as the jet propagates into the interstellar medium.  All of this materials could, in principle, have a different composition. 
In light of this, a constant value of $Y_{e}$ applied to the entire RHD domain is just an approximation. To ensure that such approximation gives reliable results, we restrict this study to the region near the jet axis by injecting photons only within the first $3^{\circ}$ relative to the jet axis, where the jet material has a high temperature and Lorentz factor. The role of mixing between materials with different $Y_{e}$ will be explored in a future work.  
\section{Results} 
In this paper we used the FLASH version 2.5 2D RHD simulation in \cite{lazzati_photospheric_2013} that is based on a 16TI progenitor \citep{woosley_progenitor_2006} in which a jet with initial Lorentz factor of 5 and \sout{an} opening angle of 10\textdegree  \hspace{0.5mm} is injected into the 16TI progenitor for 100 s and propagates out to the photosphere at $ \sim  10^{13} $ cm. The 16TI simulation in  \cite{lazzati_photospheric_2013} was performed on an adaptive mesh grid with a maximum resolution of $4 \times 10^{6}$ cm and output files were saved every $\delta t = 0.2$ s. For our MCRaT simulations to converge according to \cite{arita-escalante_optimizing_2023}, injected photons should travel through multiple RHD cells in each frame.  This can be quantified through the light crossing ratio, defined as $c\delta t / \delta r$  which, with the spatial and temporal resolutions from the 16TI simulation used here, results in a light crossing ratio as large as $\sim 1500$. 

Our methods are similar to \cite{parsotan_photospheric_2018}, with a key difference being that we inject $\sim 2\times 10^{5}$ photons for $\sim 50$ s of a non-variable jet, which excludes only a constant, low luminosity portion of the lightcurve that is not observed in nature. We also restrict photon injection to the first 3\textdegree of the jet as outlined in Section 2. We then adopt a viewing angle of $\theta_{v} < 3$\textdegree when conducting mock observations. For the electron-to-nucleon ratio we use the values $Y_{e} = 1, 0.7, 0.4, \, \text{and} \, \, 0.1$ to cover the cases of a small to large neutron component. 

\begin{figure}
\vspace*{3mm} 
\includegraphics[height=0.88\textheight]{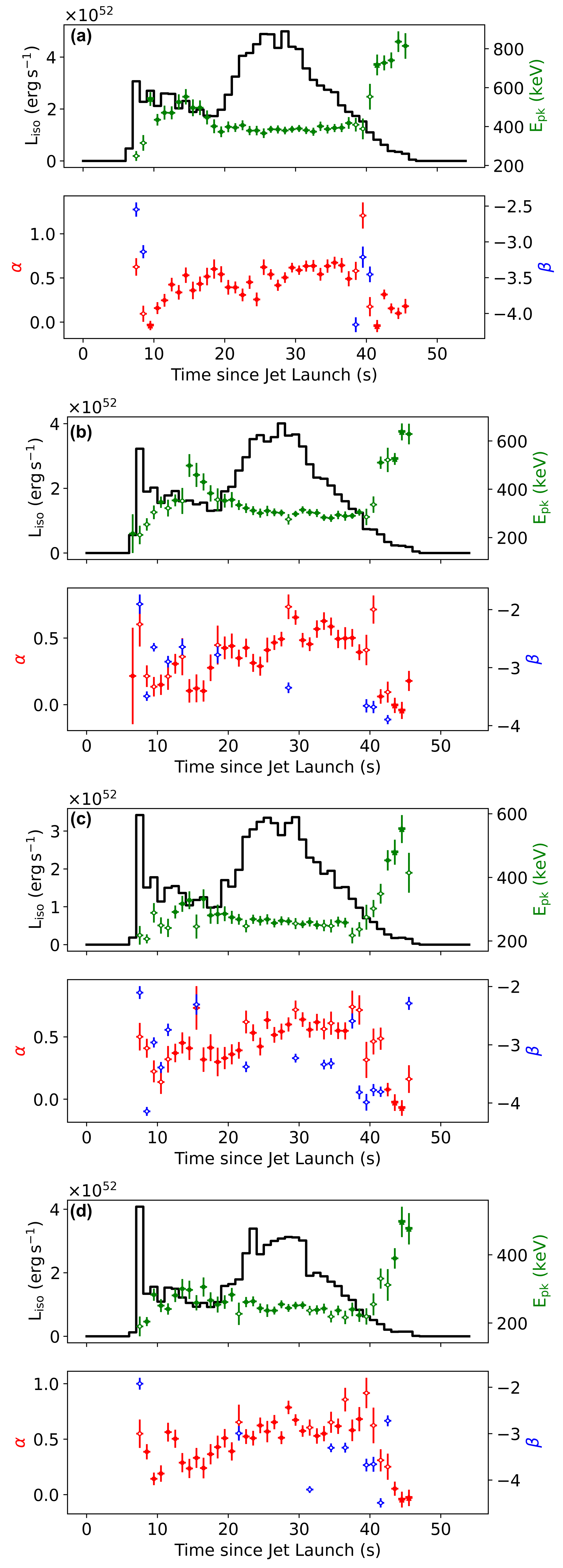}
\caption{Light curves and time resolved best fit parameters of the 4 MCRaT simulations: \textbf{(a.)} $Y_{e} = 0.1$, \textbf{(b.)} $Y_{e} = 0.4$, \textbf{(c.)} $Y_{e} = 0.7$, \textbf{(d.)} $Y_{e} = 1$.  The best fit parameter $\alpha$ is shown in red, $\beta$ is shown in blue, and $E_{pk}$ is shown in green. $\beta$ is not shown when a comptonized function provides a better fit than the Band function }
\label{fig:lightcurve}
\end{figure}

Figure \ref{fig:lightcurve} shows lightcurves obtained at a viewing angle of $\theta_{v} = 1^{\circ}$ alongside the time-resolved best fit parameters $\alpha$ and $\beta$ for the Band function (equation \ref{eq:band}), in addition to the peak energy $E_{pk} = (2 + \alpha)\, E_{0}$, for all 4 values of $Y_{e}$.  Our lightcurve from the $Y_{e} = 1$ simulation agrees well with past MCRaT results based on similar 16TI simulations \citep{lazzati_monte_2016,parsotan_photospheric_2021}, and all lightcurves show a characteristic small peak at $\sim 8$ s, with a brighter peak at $\sim 30$ s. As $Y_{e}$ is increased, the second peak dims noticeably as evident in panel \textbf{(d)} of Figure \ref{fig:lightcurve}, where the first peak is brighter than the second. 

\begin{figure}
\vspace*{3mm} 
\includegraphics[width=\columnwidth]{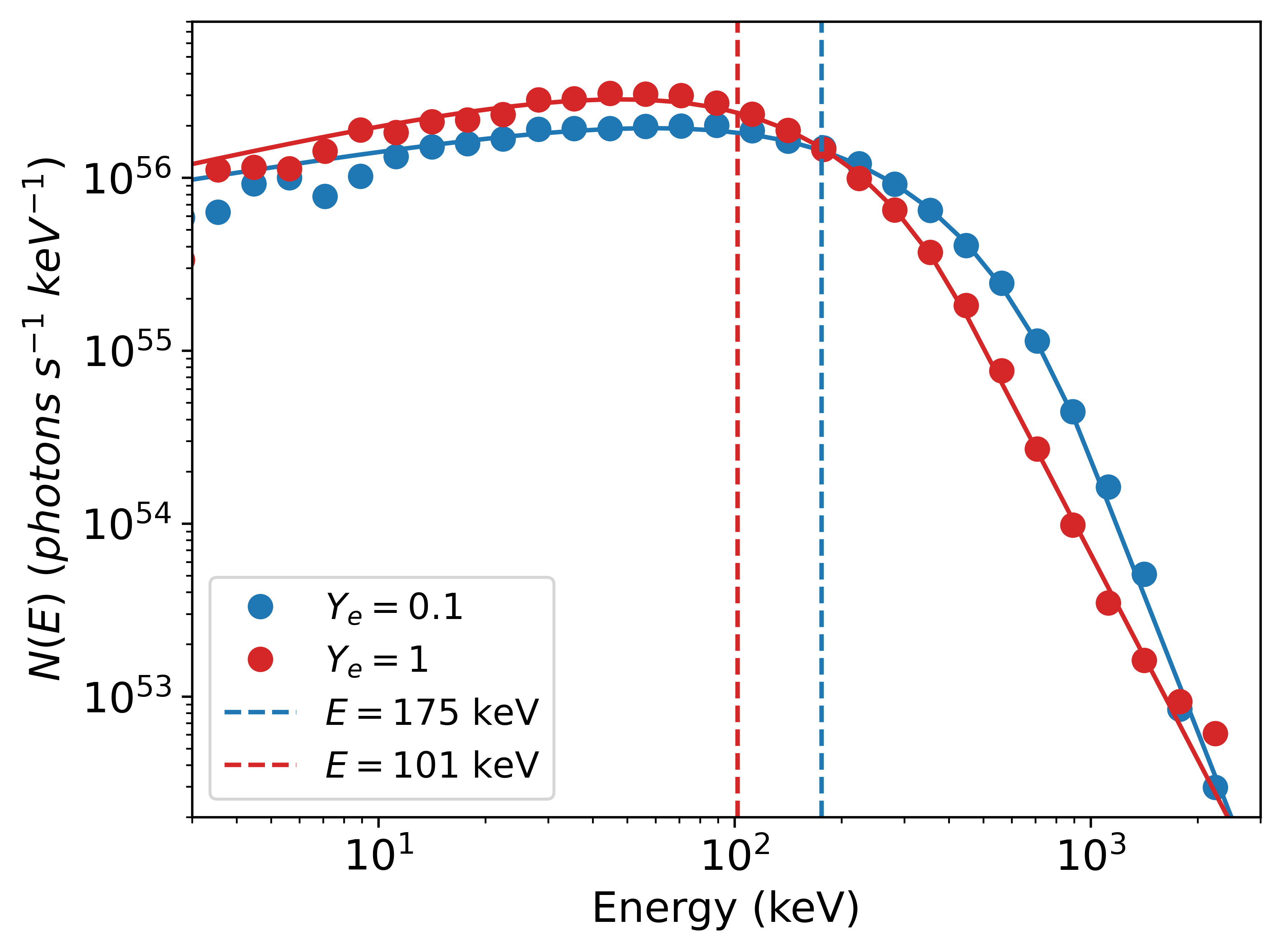}
\caption{Time-integrated spectra for MCRaT simulations with $Y_{e} = 1$, shown in red, and $Y_{e}=0.1$, shown in blue. In both cases circles show data points and the solid lines show the best fit Band functions.  The vertical dashed lines show the break energies, $E_{0}$, for both simulations. Both spectra were calculated using photons collected over the the first 40 s of the lightcurves in figure \ref{fig:lightcurve}.}
\label{fig:spectra}
\end{figure}

In Figure \ref{fig:spectra} we show time-integrated spectra obtained from photons in the $Y_{e} = 0.1$ and $Y_{e} = 1$ MCRaT simulations that have reached the final RHD frame. Both spectra in figure \ref{fig:spectra} were integrated from 0 to 40 s, corresponding to the first two peaks seen in figure \ref{fig:lightcurve}. As with Figure \ref{fig:lightcurve}, our spectra with $Y_{e} = 1 $ agrees well with past results. Here, as $Y_{e}$ is decreased, the peak energy shifts to higher frequencies as seen by the dotted lines in Figure \ref{fig:spectra}. We will look at how $Y_{e}$ affects other spectral parameters below.

\begin{figure}
\includegraphics[width=\columnwidth]{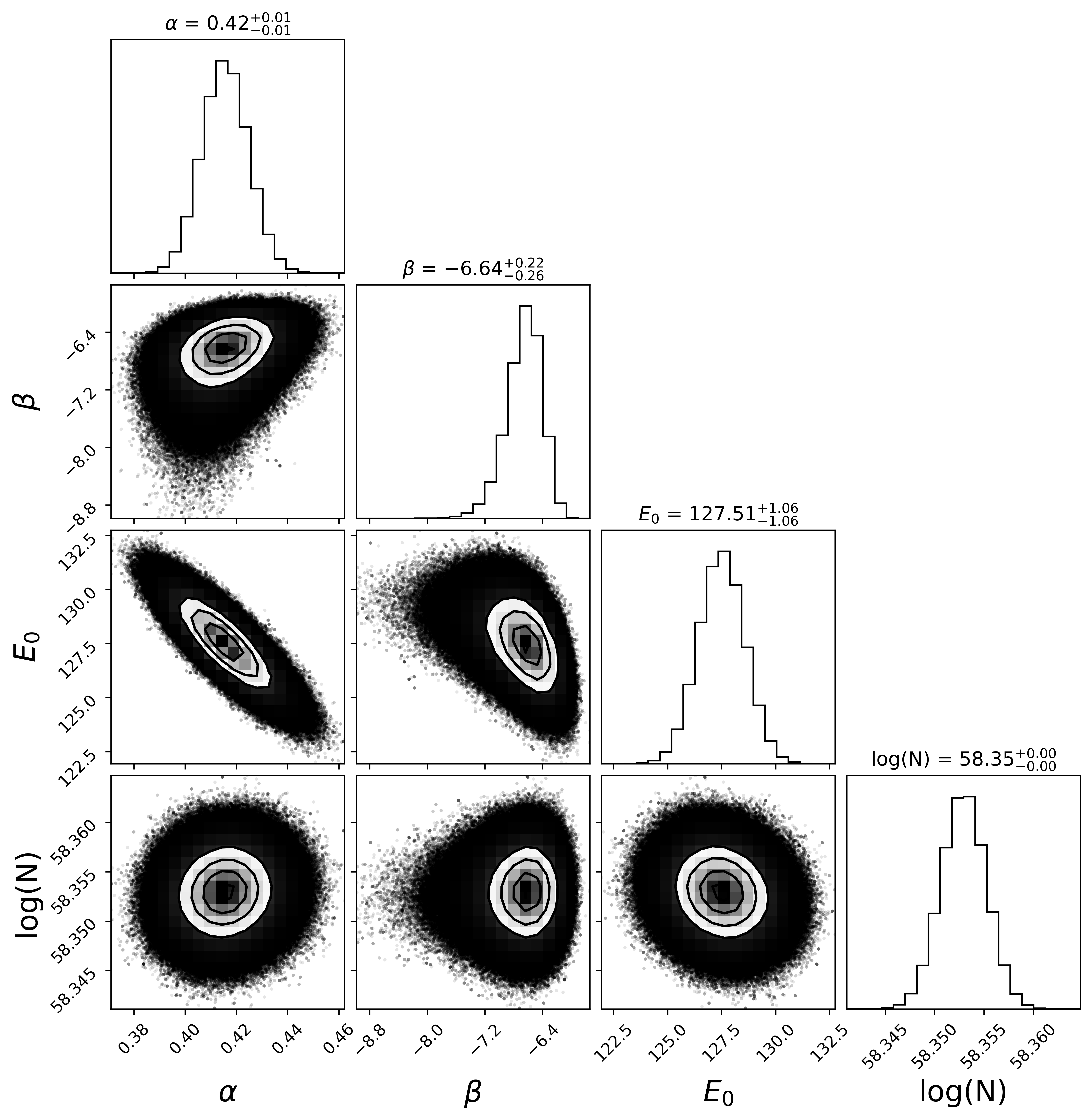}
\caption{Corner plot resulting from fitting the Band function to the spectrum from the $Y_{e} = 0.1$ simulation with a Markov-Chain Monte Carlo algorithm. The four parameters are the low energy slope $\alpha$, the high energy slope $\beta$, the break energy $E_{0}$, and the normalization parameter $N$.  This clearly shows a tight correlation between $E_0$ and $\alpha$, with less prominent correlations between all other parameters. }
\label{fig:corner}
\end{figure}

Figure \ref{fig:corner} shows a corner plot for a Band function fit to the $Y_{e}=0.1$ spectrum.  While spectral parameters in figures \ref{fig:lightcurve} and \ref{fig:spectra} where obtained via a non-linear least squares fitting algorigthm available in ProcessMCRaT, those in Figure \ref{fig:corner} were obtained by fitting a Band function to our MCRaT data with a Markov Chain Monte Carlo algorithm via emcee \citep{foreman-mackey_emcee_2013}. The parameters in Figure \ref{fig:corner} are different from those seen in Figure \ref{fig:spectra} due to the different methodologies used to obtain them.  Figure \ref{fig:corner} shows a clear correlation between $E_{0}$ and $\alpha$, while the other pairs of parameters have no notable correlations.  This strong correlation between $\alpha$ and $E_{0}$ plays a part in the evolution of Band function parameters for all four of the MCRaT simulations in this work.  

\begin{figure*}
 
\includegraphics[width=\textwidth]{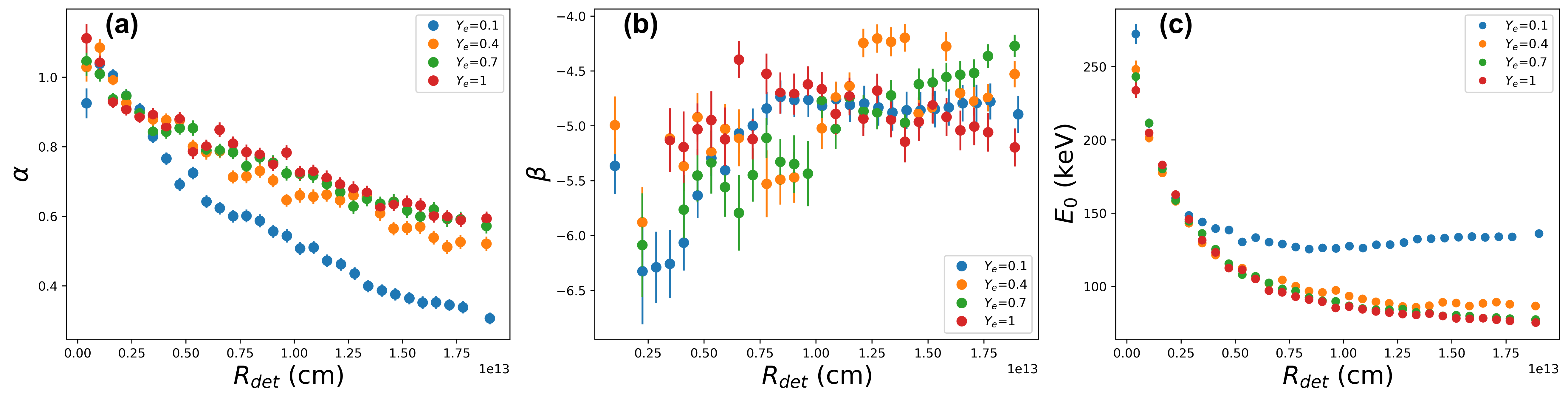}

\caption{Evolution of the Band function parameters for spectra computed from each value of $Y_{e}$: \textbf{(a)} the low energy slope $\alpha$; \textbf{(b)} the high energy slope $\beta$; and \textbf{(c)} the break energy $E_{0}$.  Each data point represents a parameter obtained from a mock observation conducted at various intermediate steps throughout the MCRaT simulation. At each step, the injected photons have some average position and a detector was placed slightly beyond that point, denoted $R_{det}$. As the simulation progresses, the position of the detector moves further away from the central engine and the Band function parameters approach their final values near the photosphere. Panels \textbf{(a)} and \textbf{(c)} show clear patterns for $\alpha$ and $E_{0}$, respectively.  Spectra obtained for all four values of $Y_{e}$ start off hot, having a high $E_{0}$, and gradually cool as the MCRaT simulations progress. $E_{0}$ obtained from the $Y_{e}=0.1$ simulation levels off sooner than for the other simulations, and so maintains a hotter spectra. This behavior is mirrored in panel \textbf{(a)}, with $\alpha$ reaching a smaller value for $Y_{e}=0.1$ than for other simulations.  Panel \textbf{(b)} shows no discernible pattern for $\beta$.}
\label{fig:band}
\end{figure*}

It is also illuminating to analyze the behaviour of the Band Function parameters as the radiation propagates with and through the outflow material. We do this by conducting a mock observation and calculating spectra for multiple intermediate times throughout the MCRaT simulation.  At each of these times, the injected photons have scattered through only a portion of the RHD simulation, and thus have some average distance from the central engine.  This distance increases as the photons propagate with the outflow until they near the photosphere. For these observations, the position of the detector is determined by the positions of the photons at a given frame. The Band function is fit to the spectrum at each time, and  Figure \ref{fig:band} shows how the Band function parameters $\alpha$, $\beta$, and $E_{0}$ vary as a function of the photons' average distance from the central engine for all values of $Y_{e}$ we consider.  As with Figure $\ref{fig:spectra}$, all parameters come from time-integrated spectra.

Panels \textbf{(a)} and \textbf{(c)} in Figure \ref{fig:band} clearly show the imprint of a neutron component on the spectral parameters of LGRBs.  All four of our MCRaT simulations start off hot near the central engine and gradually cool as the photons and outflow propagate.  Simulations with a smaller neutron component cool down more, resulting in lower peak energies.  Since $E_{0}$ and $\alpha$ are correlated (e.g. Figure \ref{fig:corner}), the low energy slope $\alpha$ mirrors this behavior, with simulations having larger neutron components displaying smaller values for $\alpha$. Panel \textbf{(b}), however, shows no clear trend.

\begin{figure}
\includegraphics[width=\columnwidth]{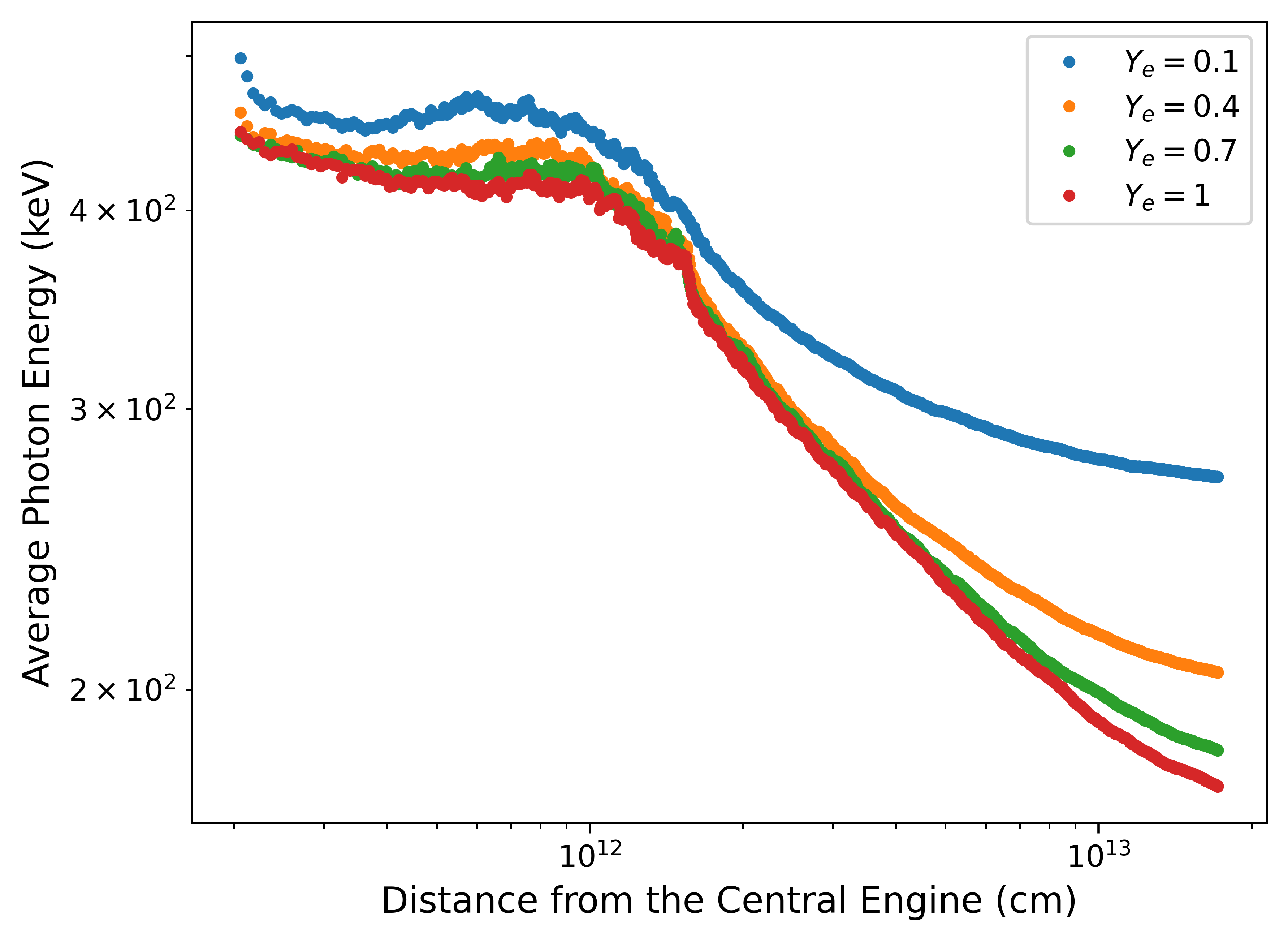}
\caption{Average photon energy computed as a function of distance from the central engine. Injected photons in all four of our simulations start of with similar energy and as the photons propagate further from the central engine photons in simulations with lower values of $Y_{e}$ begin to decouple from the jet sooner, resulting in higher energies for those simulations. The energy from the $Y_{e}=0.1$ is nearly constant after $R \sim 10^{13}$ cm, while the rest appear to be somewhat coupled to the jet by the time the photons reach the last RHD simulation frame at $R \sim 10^{13}$ cm.  }
\label{fig:avgE}
\end{figure}

\begin{figure}
\includegraphics[width=\columnwidth]{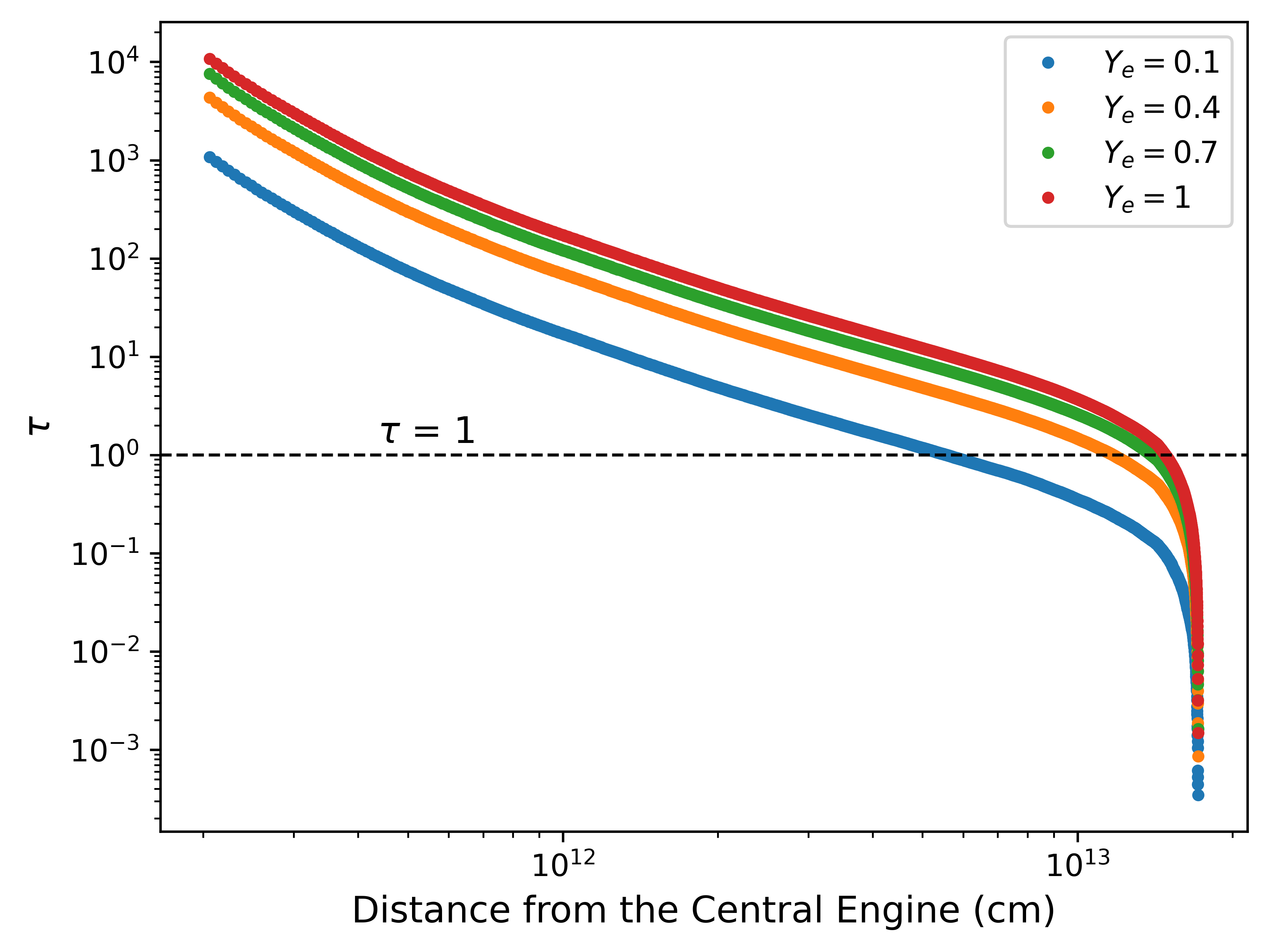}
\caption{Optical depth (Equation \ref{eq:tau}) for all four of our MCRaT simulations. Scatterings for each photon are counted, starting when they're injected near the central engine, and accumulate as they propagate out to the photosphere. Initially, $\tau \sim 10^{3} - 10^{4}$ which is high enough to ensure that the photons are described by a Planck spectrum. There is a significant drop in $\tau$ at $\sim 1.8 \times 10^{13}$ cm, which corresponds to the average photon position in the last RHD frame.  Photons that are fully decoupled from the outflow have an optical depth of $\tau \sim 1$, and the MCRaT simulations with $Y_{e} = 0.1$ and $Y_{e} = 0.4$ reach this value before the drop.  The MCRaT simulations with $Y_{e} = 0.7$ and $Y_{e} = 1$, however, reach this value right at the edge of the drop, indicating that these simulations are still somewhat coupled to the outflow.}
\label{fig:tau}
\end{figure}

Figure \ref{fig:avgE} shows the average photon energy as a function of their distance from the central engine. Figure \ref{fig:tau} similarly shows how the optical depth (equation \ref{eq:tau}) of the injected photons. In figures \ref{fig:avgE} and \ref{fig:tau}, the photon energy and number of scatterings for each photon are, respectively, calculated for every individual photons starting immediately when they're injected near the central engine. 

As stated in the Methods section, for the spectra and lightcurves from MCRaT to correspond what an observer would see from an actual burst, the photons have to decouple from the jet material.  Figure \ref{fig:tau} shows this directly. All four MCRaT simulations considered here start off with photons that have an optical depth of $10^{3}-10^{4}$.  As the photons scatter and propagate with the outflow, their optical depth slowly decreases until it reaches a sharp decay at $\sim 1.8 \times 10^{13}$ cm. While the photons in all of our MCRaT simulations reach $\tau =1$, some only do so at this sharp drop. This rapid decay is due to the sum in Equation \ref{eq:tau} only going to the last RHD simulation frame, instead of all the way out to infinity.  The fact that our MCRaT simulations with $Y_{e} = 0.7$ and $Y_{e} = 1$ only reach an optical depth of 1 when this artificial drop occurs is indicative of the fact that the photons in these simulations are still relatively coupled to the outflow. A proxy for this can be seen in Figure \ref{fig:avgE}, which shows the same cooling behavior as panel \textbf{(c)} in Figure \ref{fig:band}, with photon energies beginning to level off as they approach the photosphere.  In particular, it also shows that the photon energy for the simulation with $Y_{e} = 0.1$ has nearly leveled off while the energies for the other three simulations are still actively decreasing, indicating that the photons are still scattering with the outflow.  

\begin{figure*}
 
\includegraphics[width=\textwidth]{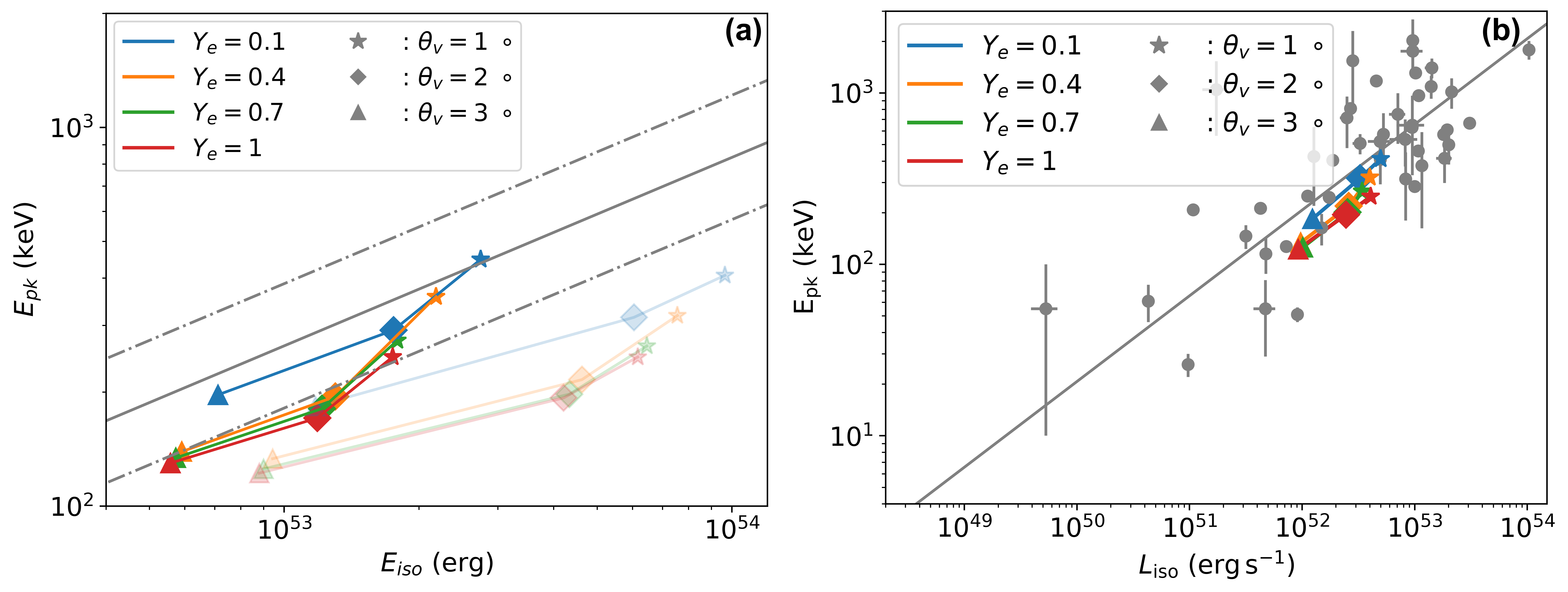}

\caption{\textbf{a.)} Amati and \textbf{b.)} Yonetoku correlations for all four values of $Y_{e}$.  To obtain multiple observations for each simulation, we conduct a mock observation at three different viewing angles.  In each figure, different shapes denote viewing angles and different colors denote different values of $Y_{e}$. In \textbf{a.)}, the solid gray line shows the Amati Relationship from \cite{tsutsui_cosmological_2009}, with the dotted gray line showing the $1\sigma$ confidence intervals. The faded colors show data obtained from the first 40 s of the lightcurves in Figure \ref{fig:lightcurve}, while the solid colors show only the first 20 s. In \textbf{b)}, the gray dots show observational data of GRBs from \cite{nava_complete_2012}, with the solid gray line showing the line of best fit. All MCRaT simulations follow the Yonetoku relation, with lower values of $Y_{e}$ corresponding to higher $E_{pk}$, $E_{iso}$, and $L_{iso}$. Similarly to past work with MCRaT, there is some strain with the Amati relation, but this strain is removed when only considering photons from the first 20 s of the jet, when it is experiencing more shocks.  Simulations with more neutrons fit both relations better, regardless of which portion of the lightcurve we consider.}
\label{fig:corAY}
\end{figure*}

In past works, MCRaT has had successes in reproducing various observational correlations of GRBs \citep{parsotan_photospheric_2018}.  Figure \ref{fig:corAY} shows the Amati and Yonetoku correlations for the four simulations considered here, with viewing angles of $\theta_{v} = 1^{\circ}, \, 2^{\circ}, \,  3^{\circ}$. The Amati relation in panel \textbf{(a)} shows two sets of points, one set corresponding to calculations using photons from the first 20 s of the lightcurves in Figure \ref{fig:lightcurve} (shown in solid colors), while the other set uses photons from the first 40 s (shown in faded colors). Panel \textbf{(b)} shows the Yonetoku relation for photons obtained only during the first 40 s. Here, we see that our simulations agree well with the Yonetoku relation, regardless of $Y_{e}$ or which portion of the lightcurve we use. With the Amati relation, we find that there is some strain when using photons from all 40 s, which is similar to results from \cite{parsotan_monte_2018}. However, we can recover the relation if we restrict ourselves to photons from the first 20 s. 

This is not an entirely new result, since MCRaT analysis of a similar simulation with a short-lived engine \citep{parsotan_photospheric_2018} yielded analogous results. Qualitatively, it is also expected that shortening the duration of the engine reduces the total burst energy (moving points to the left in the Amati plane) with only a relatively small effect on the peak photon energy, likely in the upward direction since bursts tend to have harder spectra in their early phases.



\begin{figure}
\includegraphics[width=\columnwidth]{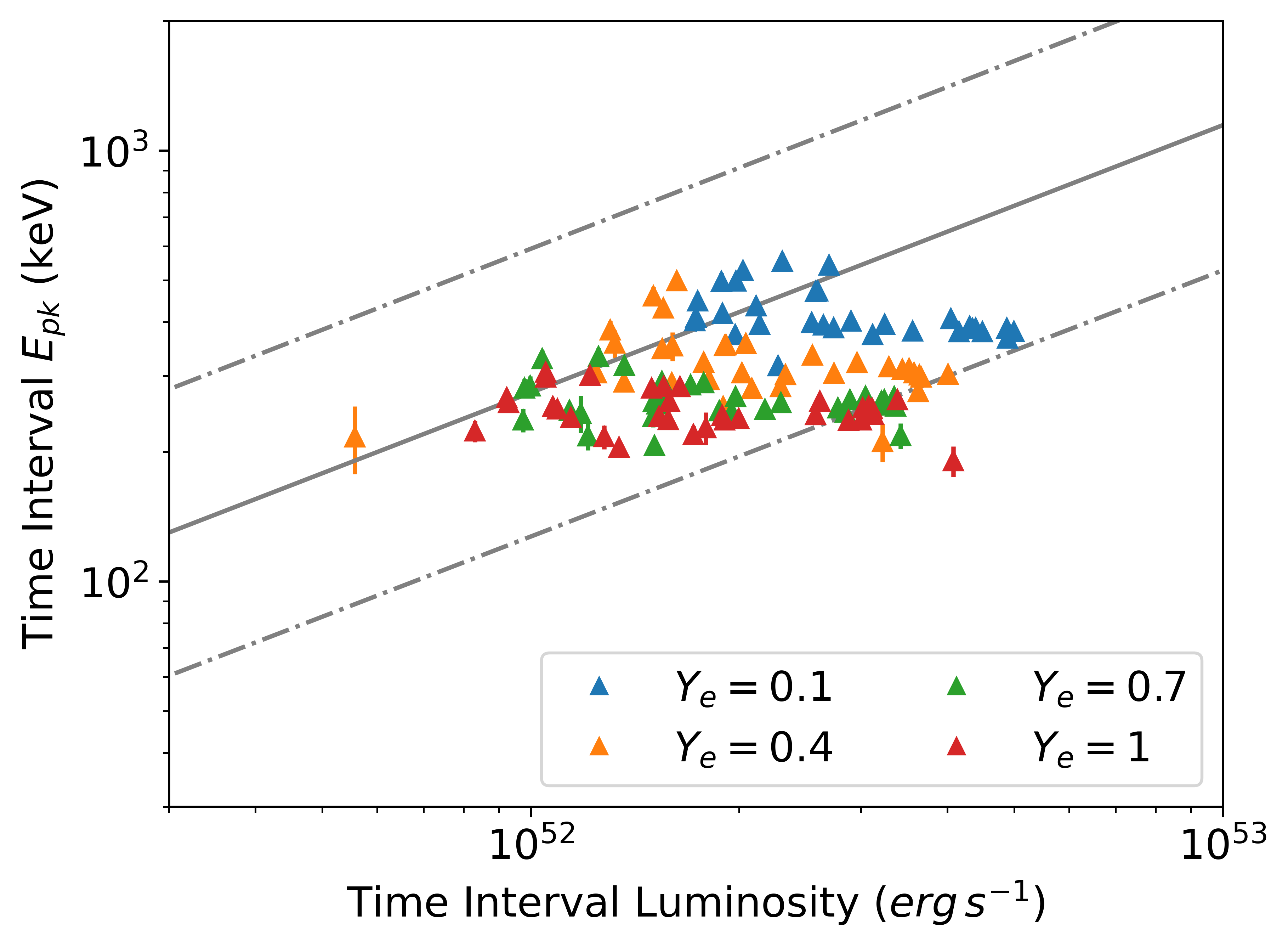}

\caption{Golenetskii relation for all values of $Y_{e}$ over the first 40 s of each burst.  Each value of $Y_{e}$ is denoted by a different color, and each point is calculated by binning the lightcurves shown in Figure \ref{fig:lightcurve} into $1$ s bins and calculating the time resolved $E_{pk}$ for each bin.  The gray solid indicates the Golenetskii relation from \cite{lu_comprehensive_2012}, with the dotted gray lines representing the $2\sigma$ intervals.  Every simulation shows good agreement with the Golenetskii relation, with smaller values of $Y_{e}$ corresponding to higher values of $E_{pk}$ and Luminosity, similar to Figure \ref{fig:corAY} } 
\label{fig:corG}
\end{figure}

Figure \ref{fig:corG} shows the Golenetskii relation for all values of $Y_{e}$. Each point is calculated by finding the luminosity and time resolved $E_{pk}$ over 1 s time bins for the first $20$ s of the lightcurves in Figure \ref{fig:lightcurve}.  As with the Yonetoku relation, we find good agreement with observations without any restrictions on $Y_{e}$ or photons.  Moreover, we find that simulations with a larger neutron component tend to push peak energies and luminosities into better agreement with all three observational correlations.

\begin{figure}
 
\includegraphics[width=\columnwidth]{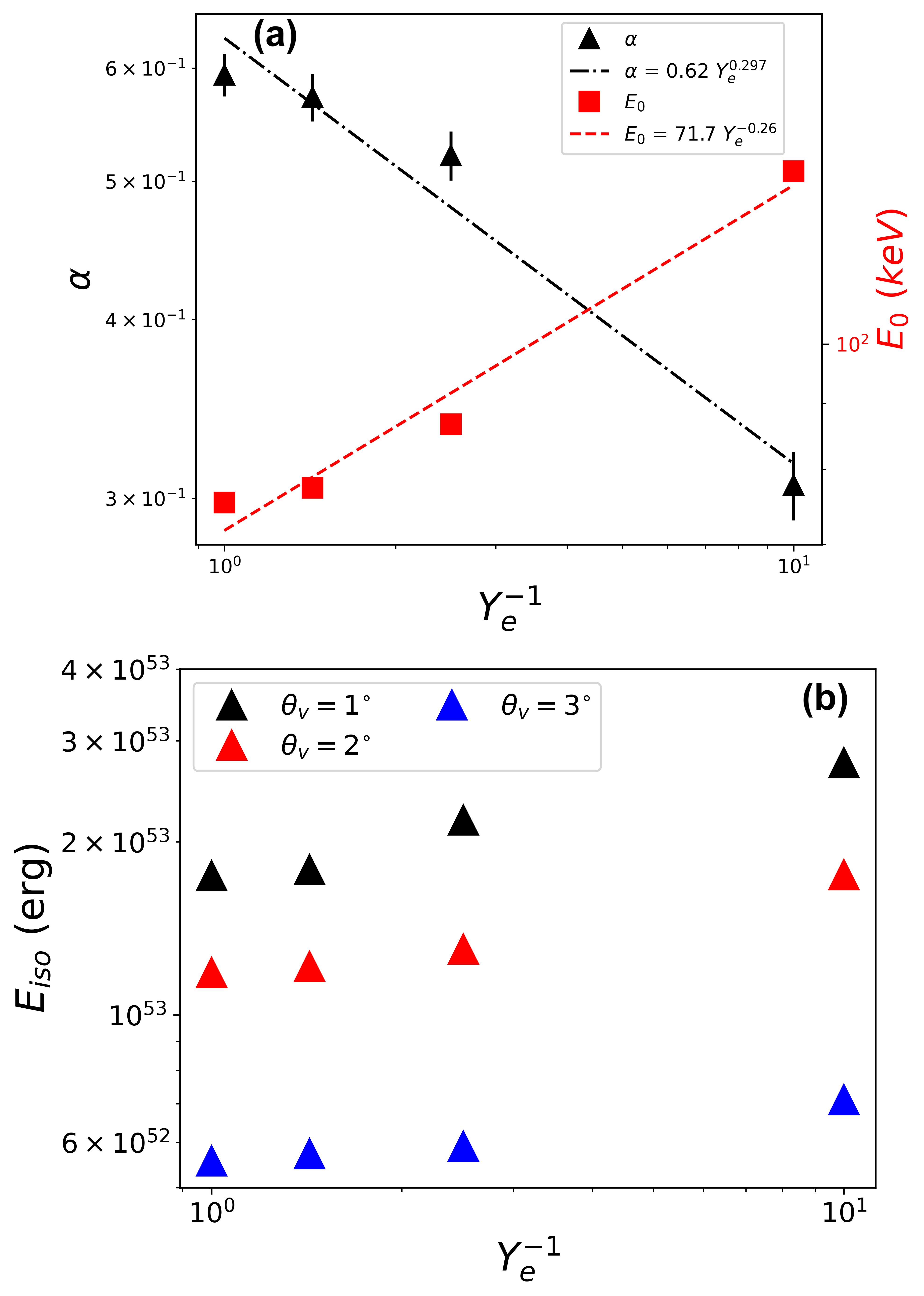}

\caption{The $Y_{e}$ effect on \textbf{(a)} the Band function parameters $\alpha$ and $E_{0}$, and on \textbf{(b)} total radiated energy. The x-axis in both panels shows $Y_{e}^{-1}$ so the size of the neutron component increases to the right. In \textbf{(a)}, the red squares show the break energy $E_{0}$ and the black triangles show the low energy slope $\alpha$, with the dashed and dashed-dotted lines showing the best fit trend lines for $E_{0}$ and $\alpha$, respectively. The break energy $E_{0}$ clearly increases as the neutron component gets larger, and $\alpha$ clearly decreases nearly symmetrically as evidenced by the $E_{0}$ slope of -0.26 and the $\alpha$ slope of 0.297.  The low energy slope $\beta$ is not shown due to a lack of a clear pattern in Figure \ref{fig:band}. In \textbf{(b)} the different colors show the isotropic energy from mock observations conducted at different viewing angles.  As the neutron component is increased, the total radiated energy is increased across multiple viewing angles.}
\label{fig:Ye}
\end{figure}

The role of the neutron component in our simulations can be summarized by plotting spectral parameters as a function of $Y_{e}$. Panel \textbf{(a)} in Figure \ref{fig:Ye} shows how the Band parameters $\alpha$ and $E_{0}$ depend on $Y_{e}$, with best-fit power laws shown as dashed and dash-dotted lines. $\beta$ is not shown due to the lack of a clear pattern in Figure \ref{fig:band}.  Neither $\alpha$ nor $E_{0}$ change very much when $Y_{e}$ is near 1.  However, as  the size of the neutron component increases, corresponding to our simulations with $Y_{e} = 0.4$ and $Y_{e} = 0.1$, the spectral parameters begin to change more dramatically.  This is consistent with Figures \ref{fig:avgE} and \ref{fig:tau} showing that simulations with a small neutron component are still somewhat coupled to the outflow.  Had the injected photons been able to scatter for longer, it is likely changes would be more consistent across the range of $Y_{e}$ considered here.  Furthermore, the nearly symmetric slopes of trend lines in panel \textbf{(a)} are consistent with the strong correlation between $\alpha$ and $E_{0}$ on display in Figure \ref{fig:corner}. Additionally, as suggested by Figures \ref{fig:corAY} and \ref{fig:corG}, panel \textbf{(b)} in Figure \ref{fig:Ye} shows that the radiative efficiency increases as the size of the neutron component is increased, and that this effect isn't dependent on viewing angle for the range considered here.

\section{Summary and Discussion} 

In this paper we present results from a series of MCRaT radiative transfer simulations that probe the role that a neutron component in the outflow has on the radiation produced in a LGRB. Varying the density of the input RHD simulation controls the size of the neutron component via the lepton density in Equation \ref{eq:lambda}, which in turn changes how the photons interact with the outflow until they reach the photosphere.

Observables, such as spectra and lightcurves, can be produced with the results of our MCRaT simulations. Our  $Y_{e}=1$ lightcurve, and the associated time-resolved spectral parameters, show good agreement with past works using similar 16TI RHD simulations  (e.g. \cite{parsotan_photospheric_2021}).  We likewise find good agreement between our $Y_{e} = 1$ time-integrated spectra and those seen in the same paper.

We find clear patterns in the spectral parameters as we vary $Y_{e}$. In particular, the break energy $E_{0}$ (and thus the corresponding peak energy $E_{pk} = [2 + \alpha] E_{0}$) is shifted to higher energies as $Y_{e}$ decreases (and the size of the neutron component increases). A power-law fit to $E_{0}$ as a function of $Y_{e}^{-1}$ ($E_0\propto Y_e^{\zeta}$) yields an index $\zeta=-0.26$. This behavior is consistent with how the radiation in each of our MCRaT simulations decouple from the outflow. Our simulations with $Y_{e} = 1$ and $Y_{e} = 0.7$ are still relatively coupled to electrons in the outflow and so the photons are still appreciably cooling when they reach the last frame of the RHD simulation, resulting in a relatively weak power-law index. We also find that $\alpha$ obtained from simulations with a smaller $Y_{e}$ is consistent with a less thermal spectrum than when $Y_{e}$ is larger, and that this behavior is likely due to a strong correlation between $E_{0}$ and $\alpha$. This is supported corresponding power law for $\alpha (Y_{e}^{-1})$, which is 0.297. In contrast to the other parameters, $\beta$ has no clear trend, possibly due to the fact that the high energy tail of the spectrum forms relatively close to the photosphere compared to the lower frequency parts of the spectrum, which are characterized by $\alpha$ and $E_{0}$.  

We also show how radiation evolves from the injected blackbody to the observed Band-type spectra by conducting mock observations, and calculating spectra, using photons before they have finished scattering through the final RHD simulation frame. This shows that all parameters start off more or less equal across all our simulations, and at some point they begin to diverge until they settle to their final values near the photosphere.  In particular, $E_{0}$ starts off relatively high and decreases gradually as the injected photons propagate through and with the outflow. The low frequency index $\alpha$ mirrors this behavior, probably due to their strong correlation.

Similar behaviour is observed when we track the optical depth and average energy of the injected photons, beginning immediately after injection until they finish scattering.  Both quantities start out high, indicating that that the photons are injected into a hot and dense outflow, and so are well-described by the blackbody spectrum.  We see a gradual decoupling of the photons from the outflow, which mirrors the behaviour of the spectral parameters.

Finally, we check our simulations against the observational correlations of Amati, Yontetoku, and Golenetskii \citep{amati_intrinsic_2002,yonetoku_gamma-ray_2004,golenetskii_correlation_1983}, and find good agreement with all three, regardless of $Y_{e}$.  This agrees well with past work with MCRaT \citep{parsotan_photospheric_2018}.  However, given the maximum injection angle of $3^{\circ}$, we are limited to the number of observations we can make.  Interestingly, while all of our simulations fit these correlations nicely, those with a larger neutron component tend to lie closer to the trend lines than those with a smaller neutron component.

Generally, these results are very promising as they provide a mechanism for increasing the peak energy predicted by photospheric models of GRB prompt emission. While there is no consensus on the neutron content of GRB outflows, their presence in both core collapse supernov\ae{} and binary neutron star mergers suggests that peak energies are at least somewhat higher than seen in past works with MCRaT. The corresponding increase in total radiated energy (which is inevitable since the number of photons is conserved in a pure scattering process) increases radiative efficiency and brings the MCRaT predictions to better agreement with observational correlations.  Both of these results can be interpreted by considering a baryon-loaded LGRB outflow:  when the outflow is produced near the central engine, it is hot and dense and thus produces blackbody radiation.  The outflow is subsequently heated via shocks as it bores its way through the stellar envelope.  Eventually the outflow will clear the envelope and begin to cool while its internal energy is  converted to bulk kinetic energy.  Thus, the initially hot blackbody radiation also cools as it gradually decouples from the matter component of the outflow.  When there is a neutron component in the outflow, radiation will decouple sooner and will thus carry with it a signature of the outflow from when it had converted less of its internal energy into bulk kinetic energy, thereby resulting in the observed increase in radiative efficiency. 

An important consideration of the material component of GRB outflows, not treated here, is that of mixing.  The jet, cocoon, and stellar envelope could all have different neutron components, and mixing between these could thus modify observables.  This effect would likely be more prominent at larger viewing angles where mixing is more prominent.  Furthermore, the methods discussed here could naturally be extended to sGRB simulations emerging from binary neutron star mergers.  Both of these considerations will be explored in future works.


\begin{acknowledgments}
N. W. and D.L. acknowledge support from
NSF grant AST-1907955
\end{acknowledgments}

%

\vspace{5mm}
\facilities{Resources supporting this work were provided by the NASA High-End Computing (HEC) Program through the NASA Advanced Supercomputing (NAS) Division at Ames Research Center.}





\bibliography{manuscript}
\bibliographystyle{aasjournal}



\end{document}